\documentclass[journal,12pt,onecolumn,draftclsnofoot,]{IEEEtran}

\IEEEoverridecommandlockouts
\usepackage{cite}
\usepackage{amsmath,amssymb,amsfonts}
\usepackage{algorithmic}
\usepackage{graphicx}
\usepackage{textcomp}
\usepackage{xcolor}
\def\BibTeX{{\rm B\kern-.05em{\sc i\kern-.025em b}\kern-.08em T\kern-.1667em\lower.7ex\hbox{E}\kern-.125emX}}
\usepackage[prependcaption,disable]{todonotes}
\presetkeys{todonotes}{inline}{}
\usepackage[utf8]{inputenc}
\usepackage[T1]{fontenc}
\usepackage{grffile}
\graphicspath{{figures/}}
\usepackage{subcaption}
\usepackage{bbm}
\usepackage{url}
\usepackage{etoolbox}
\newtoggle{isit}
\iftoggle{isit}{}{\usepackage{hyperref}}%
\usepackage[shortlabels]{enumitem}
\usepackage{algorithm}
\usepackage{algorithmic}  

\DeclareMathOperator*{\minimize}{minimize}%
\newcommand{\defeq}{\triangleq}
\newcommand{\reals}{\mathbb{R}}
\renewcommand{\Pr}{\operatorname{\mathbb P}}
\newcommand{\E}{\operatorname{\mathbb E}}
\newcommand{\law}{\operatorname{\mathcal L}}

\newcommand{\ind}[2]{\left(#1\colon #2\right)} 
\newcommand{\cov}{\operatorname{Cov}}

\newtheorem{exam}{Example}
\newtheorem{definition}{Definition}
\newtheorem{remark}{Remark}
\newtheorem{prop}{Proposition}

\newcommand{\T}{^{*}}
\newcommand{\indicator}[1]{I_{#1}}
\newcommand{\erPrior}{\pi^0_{\mathsf{ER}}}%
\newcommand{\figWidth}{0.48\textwidth}%
\newcommand{\sigmat}{\nu}
\newcommand{\sigmab}{\sigma}
\newcommand{\meanInit}{\mu}
\newcommand{\covInit}{Q}
\newcommand{\edgeFNR}{P^-}%
\newcommand{\edgeFPR}{P^+}%
\newcommand{\fnrNet}{\epsilon^-}%
\newcommand{\fprNet}{\epsilon^+}%
\newcommand{\Best}{\widehat B}%
\newcommand{\uiuc}{University of Illinois Urbana–Champaign}

\renewcommand{\hat}{\widehat}
\newcommand{\lr}{\mathsf{LR}}
\newcommand{\er}{Erd\H{o}s–R\'{e}nyi}

\DeclareMathOperator{\Tr}{Tr}

\begin{document}
\listoftodos

\title{Lower Bounds on Information Requirements for Causal Network Inference\\
  \thanks{This material is based upon work supported by the National Science Foundation under Grant No.\ CCF 19-00636. A portion of this work appeared in ISIT 2021.}}

\author{%
  \IEEEauthorblockN{Xiaohan Kang and Bruce Hajek,}
  \IEEEauthorblockA{\uiuc\\
    Electrical and Computer Engineering and Coordinated Science Laboratory\\
    Urbana, Illinois\\
    Email: xkang515@gmail.com, b-hajek@illinois.edu}
}

\maketitle

\begin{abstract}
  Recovery of the causal structure of dynamic networks from noisy measurements has long been a problem of interest across many areas of science and engineering.  Many algorithms have been proposed, but there is little work that compares the performance of the algorithms to converse bounds in a non-asymptotic setting.  As a step to address this problem, this paper gives lower bounds on the error probability for causal network support recovery in a linear Gaussian setting.  The bounds are based on Monte Carlo estimation of receiver operating characteristic (ROC) curves based on likelihood ratio samples assuming side information is available.  The estimated ROC curves and curves obtained through the use of Bhattacharyya coefficients or Kullback--Leibler divergences are also compared.
\end{abstract}


\section{Introduction}
Causal networks refer to directed graphs representing the causal relationships among a number of entities, and the inference of sparse large-scale causal networks is of great importance in many scientific, engineering, and medical fields.  For example, the study of gene regulatory networks in biology concerns the causal interactions among genes and is vital for finding pathways of biological functions \cite{Alon06,MarbachPrillSchaffter10}.  Because of the scale of these networks, inference often cannot be carried out for specific ordered pairs of the vertices without significant prior knowledge about the networks.  Instead, it is desirable to infer the sparse structure from observations on all the vertices.  Time-series observations are especially useful due to the nature of causality.  The problem of causal network inference is then typically formulated as a sparse support recovery problem from time-series vertex data.

Numerous algorithms have been applied to the problem of causal network inference, and their performances have been evaluated using both generative models with ground truths and real data with putative truths (see, e.g., \cite{MarbachCostelloKuffner12} for gene regulatory network reconstruction), but there is little work that studies the theoretical converse bounds of the minimum information requirements.  The work \cite{SunTaylorBollt15} lays a theoretical foundation for causal network inference by studying a general dynamic Markovian model, and proposes the oCSE algorithm which is shown to find the causal structure of the network when the exact causation entropy information is available.  However, such information is often unavailable due to the limited amount of data and noise in the observations.

Motivated by \cite{SunTaylorBollt15}, as a first step to understand the fundamental information requirements, we study the linear discrete stochastic network in \cite{SunTaylorBollt15} as a special case of more general models.  Unlike \cite{SunTaylorBollt15}, we consider observation noise on the time-series measurements.

The main contribution of this paper is to provide numerical lower bounds on the information requirements for causal network inference in the form of lower bounds for the optimal tradeoff curve between false negative rate vs. false positive rate for the detection of network edges.  Equivalently, we provide an upper bound on the receiver operating characteristic curve (ROC) giving the optimal correct edge detection rate vs. rate of false positives.  First, it is shown that network-level error ratios are linear combinations of error probabilities associated with binary hypothesis testing (BHT) problems associated with individual edges (Proposition~\ref{prop:network-edge}).  Second, a detailed description of the signed \er\ (ER) networks model defined in \cite{SunTaylorBollt15} is given, including a discussion of scaling to a specified spectral radius.  Third, a genie is introduced providing information about all but one edge in the network, leading to a tractable bound on the probability of correct detection of the remaining edge.  This allows for generation of likelihood ratio samples which can be used to provide accurate estimates of ROC curves \cite{HajekKang25}.

A secondary contribution of the paper focuses on the difference between the optimal ROC curve and upper bounds on the optimal ROC curve implied by popular information theoretic measures -- namely, the Kullback--Leibler divergence and Bhattacharyya coefficients.  Here both the ROC curve and the information measures are based on all but one side information.  This part of the paper investigates the tightness of those information theoretic bounds.  In future applications it may be possible to compute such bounds but not the actual ROC curves; our motivation is to provide some idea about the tightness of the information theoretic bounds in the context of causal recovery.

Problems similar to the causal network inference in this paper have been studied in various settings, but nothing on converse bounds is known for a non-asymptotic regime.  In a linear system identification (i.e., a vector autoregressive model) setting, algorithms for this problem were recently studied in \cite{SimchowitzManiaTu18}, with sparsity constraint in \cite{FattahiSojoudi18a,FattahiMatniSojoudi19}, with observation noise in \cite{OymakOzay19}, and in a closed-loop setting in \cite{LaleAzizzadenesheliHassibi20}, and in both discrete-time and continuous-time settings in \cite{BentoIbrahimiMontanari10}.  Notably, the mutual incoherence property (see \cite{Fuchs05,Tropp06,ZhaoYu06,MeinshausenBuhlmann06,Wainwright09}) is often used providing performance guarantees for particular algorithms.  Lower bounds for probability of exact recovery in asymptotic settings have been studied in \cite{JedraProutiere19,BentoIbrahimiMontanari11,BentoIbrahimi14}.  The causal inference problem is also closely related to compressed sensing, but unlike traditional compressed sensing the data serves as the design matrix; the time series of system states and the underlying sparse signal is the weighted adjacency matrix $A$.

The organization of this paper is as follows.  Section~\ref{sec:model} introduces the dynamic system model of the causal network inference problem and performance metrics and discusses the role of the spectral radius of the weighted adjacency matrix.  Section~\ref{sec:network-topology} describes the network topology model adopted in the paper.  Section~\ref{sec:genie_bounds} describes the optimal decision rule based on likelihood ratios and introduces the notion of all but one side information to provide a tractable upper bound on ROC curves.  Section~\ref{sec:computing_bnds} describes two methods for computing likelihood ratio samples under the all but one side information.  Section~\ref{sec:num} shows the numerical comparison of the upper bound (based on the likelihood ratio samples) and two representative algorithms.  Section~\ref{sec:bounds_casual-recov-info-measures} gives the comparison of the optimal ROC curves (given the genie side-information) to upper bounds on such curves computed using the information theoretic measures (KL divergence and Bhattacharyya coefficients).  Future work is discussed in Section~\ref{sec:discussion}.  Appendix~\ref{app:info_bnds} presents background material about use of the information measures to provide bounds on ROC curves.
 
The computer code for all figures in this paper can be found at \cite{Kang21}.

\section{Model}
\label{sec:model}
\subsection{System dynamics}
Let $n$ be the number of network vertices and $A \in\reals^{n\times n}$ be the random weighted adjacency matrix of the network with a prior distribution $\pi^0$.  Let $X(t)$ be an $n$-dimensional random row vector representing the system state at time $t\in\{0, 1, 2, \dots, T\}$.  Assume $X(0) \sim \mathcal N(\meanInit, \covInit)$ for some covariance matrix $Q$ and
\begin{align}
  \label{eq:state_evolution}
  X(t) = X(t - 1)A + W(t),\quad t = 1, 2, \dots, T
\end{align}
where $W(t)\sim\mathcal N(0, \sigmab^2I)$ are independent driving noises with variance parameter $\sigmab^2$.  The noisy observations are
\[Y(t) = X(t)+Z(t),\quad t = 0, 1, \dots, T,\]
where $Z(t)\sim\mathcal N(0, \sigmat^2I)$ are observation noises with variance $\sigmat^2$.  The observations $Y = (Y(0), Y(1), \dots, Y(T))\in\reals^{n(T+1)}$ are jointly Gaussian given $A$.  The goal is to recover the support matrix $B$ from the observations $Y$, where $B$ is defined by $B_{ij} = 0$ if $A_{ij} = 0$ and $B_{ij} = 1$ if $A_{ij}\neq 0$.  This setting is the same as the linear discrete stochastic network dynamics in \cite{SunTaylorBollt15} and the discrete-time model in \cite{BentoIbrahimiMontanari10}.  However, the theoretical results in \cite{SunTaylorBollt15} and \cite{BentoIbrahimiMontanari10} do not consider observation noise.

\subsection{Performance metrics}
In this section we define the performance metrics of the network inference problem, and relate them to error probabilities for testing hypotheses about the existence of individual edges.

We first define the network-level error ratios.  Let $\Best \colon\reals^{n(T+1)}\to\{0, 1\}^{n\times n}$ be a support matrix estimator based on the observation $Y$.  Let $\indicator{\{\cdot\}}$ be the indicator function.  On the network level, following \cite{SunTaylorBollt15}, we define the false negative ratio $\fnrNet$ and the false positive ratio $\fprNet$ for a given network prior $\pi^0$ and an estimator $\Best$ by
\begin{equation}
  \label{eq:fnr}
  \fnrNet \defeq \frac{\E\sum_{i, j}\indicator{\{A_{ij}\neq 0, \Best_{ij}(Y) = 0\}}}{\E\sum_{i, j}\indicator{\{A_{ij}\neq 0\}}},
\end{equation}
\begin{equation}
  \label{eq:fpr}
  \fprNet \defeq \frac{\E\sum_{i, j}\indicator{\{A_{ij} = 0, \Best_{ij}(Y) = 1\}}}{\E\sum_{i, j}\indicator{\{A_{ij} = 0\}}},
\end{equation}
provided the denominators are positive.  Here the summations are over all ordered pairs, including the self-pairs.

Now we define the edge-level error probabilities.  For an ordered pair $(i, j)$ given the prior $\pi^0$ on $A$ and an estimator $\Best$, the recovery of $B_{ij}$ is a BHT problem with the probability of miss and the probability of false alarm given by
\begin{equation}
  \label{eq:fnr-edge}
  \edgeFNR_{ij} \defeq\Pr(\Best_{ij}(Y) = 0\mid B_{ij} = 1)
\end{equation}
and
\begin{equation}
  \label{eq:fpr-edge}
  \edgeFPR_{ij}\defeq\Pr(\Best_{ij}(Y) = 1\mid B_{ij} = 0).
\end{equation}
\begin{prop}
  \label{prop:network-edge}
  The network-level error ratios are convex combinations of the edge-level error probabilities:
  \[\fnrNet = \sum_{i, j}\edgeFNR_{ij}w_{ij}^-,\quad\fprNet = \sum_{i, j}\edgeFPR_{ij}w_{ij}^+,\]
  where
  \[w_{ij}^- \defeq\frac{\Pr\{A_{ij}\neq 0\}}{\sum_{k, l}\Pr\{A_{kl}\neq 0\}},\quad w_{ij}^+\defeq \frac{\Pr\{A_{ij} = 0\}}{\sum_{k, l}\Pr\{A_{kl} = 0\}}.\]
\end{prop}
The proof of Proposition~\ref{prop:network-edge} follows immediately by exchanging the summation and expectation in the numerators and the denominators in \eqref{eq:fnr} and \eqref{eq:fpr}.  Proposition~\ref{prop:network-edge} implies in order to study the network-level error ratios it suffices to study the edge-level error probabilities.

\begin{remark}
  The quantities $\fnrNet$, $\fprNet$, $w_{ij}^-$, and $w_{ij}^+$ can be interpreted as limits, assuming the number of instances of the support recovery problem converges to infinity.  First, $\fnrNet$ is the limiting ratio of the number of false negatives (edges in the ground truth that are missed in the prediction) to the total number of edges in the ground truth.  Similarly, $\fprNet$ is the limiting ratio of the number of false positives (predicted edges that are not in the ground truth) to the total number of ordered pairs with no edges.  Likewise, the weight $w_{ij}^-$ is the limiting fraction of edges that appear on the ordered pair $(i, j)$ out of all edges, and $w_{ij}^+$ is the limiting fraction of non-edges on $(i, j)$ that appear out of all non-edges.
\end{remark}
\begin{remark}
  While one can alternatively define $\fnrNet$ and $\fprNet$ in \eqref{eq:fnr} and \eqref{eq:fpr} by taking the expectation of the ratios rather than the ratios of the expectations, the presented definitions do not get overly dominated by the variation of the denominators, and the denominators of the stochastic ratios might even be zero.  In \cite{SunTaylorBollt15} the two quantities were originally defined for a pair of true and predicted networks.
\end{remark}
\begin{remark}
  \label{rem:symm}
  The weights $w_{ij}^-$'s and $w_{ij}^+$'s are determined by the prior $\pi^0$.  If the network prior $\pi^0$ is symmetric in the sense a) it is invariant under vertex permutation; and b) $\Pr\{A_{11} = 0\} = \Pr\{A_{12} = 0\}\in(0, 1)$, then $w_{ij}^- = w_{ij}^+ = \frac 1{n^2}$.
\end{remark}
\begin{remark}
  Note \eqref{eq:fnr} and \eqref{eq:fpr} weigh the self-edges and the other edges equally, whereas they could be weighted differently, or self-edges could be excluded.
\end{remark}

\subsection{On the spectral radius of $A$ and the initial covariance matrix $Q$}

Let $\rho(A)$ denote the spectral radius of $A$, defined by $\rho(A) \defeq\max\{|\lambda_1|, \ldots , |\lambda_n|\}$ where $\lambda_1, \ldots , \lambda_n$ are the eigenvalues of $A.$  For large $T$ there is a qualitative difference between stable systems, for which $\rho(A)<1,$ and unstable systems with $\rho(A)\geq 1.$  If the system is unstable then $E[\|X(t)\|^2] \to \infty$ as $t\to \infty$ so that if $\nu^2$ is not time dependent then the observation noise has much less effect for larger $t.$  Also, the eigenmodes (eigenvectors and generalized eigenvectors) of $A$ corresponding to the largest magnitude eigenvalues have an exaggerated influence.  In many applications the system dynamics have some nonlinearity that keeps the state bounded.  In using the linear Gaussian model to approximate such systems it is reasonable to select $A$ so that $\rho(A) <1.$

In many applications it makes sense to begin observing the system in steady state.  That might be the only option or else the number of observations could be limited and it is deemed better to start taking observations when the system has approximately reached steady state.  That corresponds to assuming that the initial covariance matrix $Q$ is equal to the steady state covariance, given by $Q = A^TQA + \sigmab^2I$, which is well defined and finite if and only if $\rho(A) < 1.$

\section{Network topology model}
\label{sec:network-topology}
This section presents the random network topology model used in \cite{SunTaylorBollt15}.  Roughly speaking, it is a signed \er\ graph with equal edge magnitudes scaled in order to control the spectral radius.

\begin{definition}
  \label{def:er}
  Let $n\in\mathbb N$, $p\in[0, 1]$ and let $V$ be a nonnegative random variable to be specified.  A \emph{signed \er\ network} with $n$ vertices, edge probability $p$, and scale factor $V$ is defined as follows.  Let $B\in\{0, 1\}^{n\times n}$ be an $n\times n$ matrix with entries being independent Bernoulli random variables with parameter $p.$ Matrix $B$ indicates which ordered pairs of vertices have a corresponding edge (i.e. $B$ is the adjacency matrix of a directed \er\ graph with possible self edges).  Let $R\in\{-1, 1\}^{n\times n}$ be an $n\times n$ matrix with entries being independent Rademacher random variables (i.e. $1$ or $-1$ with equal probability) indicating the potential signs of the edges.  Then a \emph{signed \er\ network} with $n$ vertices, edge probability $p$, and scale factor $V$ is given by $A=VB \circ R$, where $\circ$ denotes the Hadamard (entrywise) product.
\end{definition}
We shall next describe two options for the choice of $V.$

\begin{figure}[htb]
  \centering
  \begin{tabular}{c @{\qquad} c }
    \includegraphics[width=.4\linewidth]{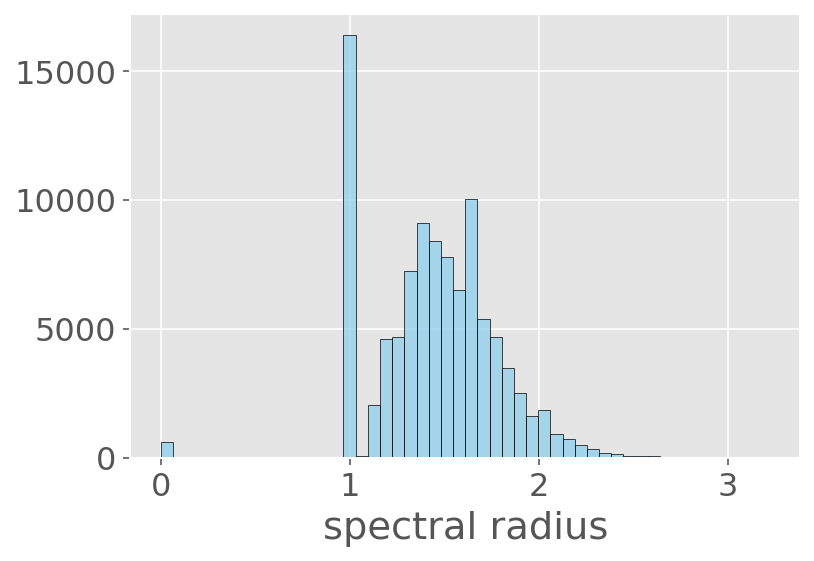} &
    \includegraphics[width=.4\linewidth]{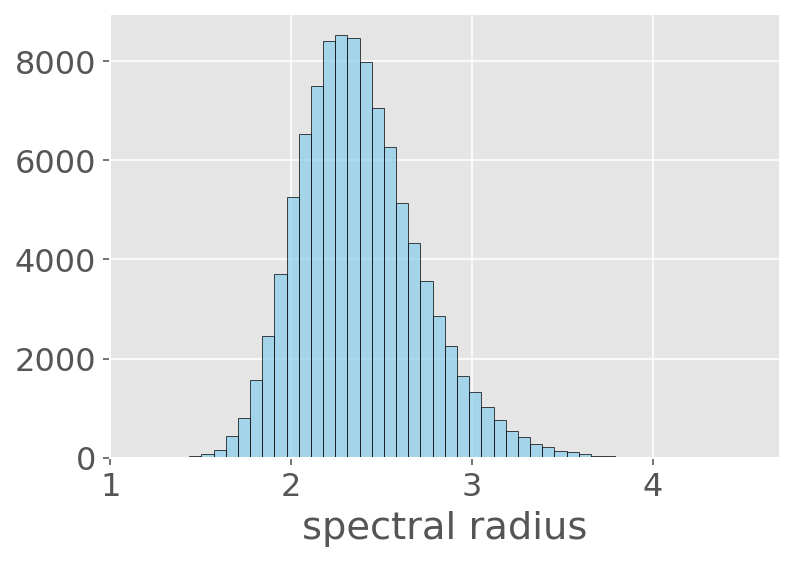} \\
    \small (a) $n=10, p=0.2.$ & \small (b) $n=10, p=0.5$  \\
        \includegraphics[width=.4\linewidth]{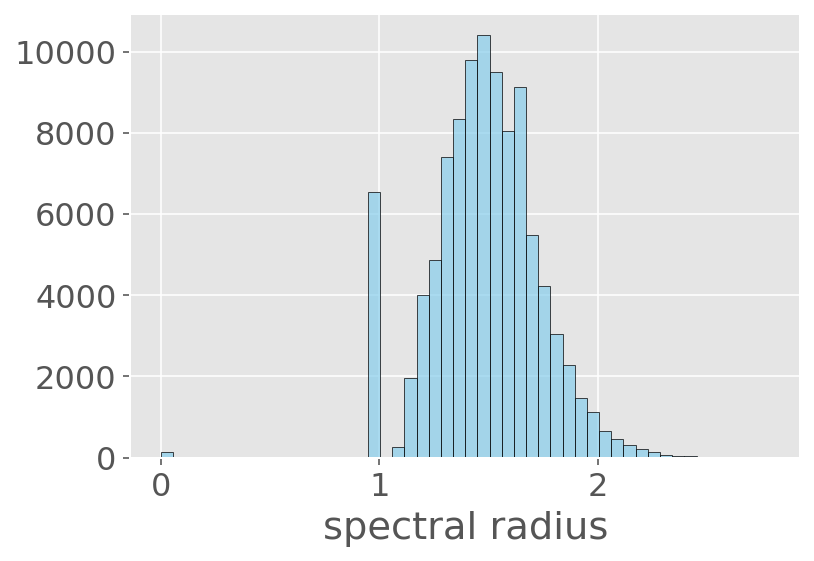} &
    \includegraphics[width=.4\linewidth]{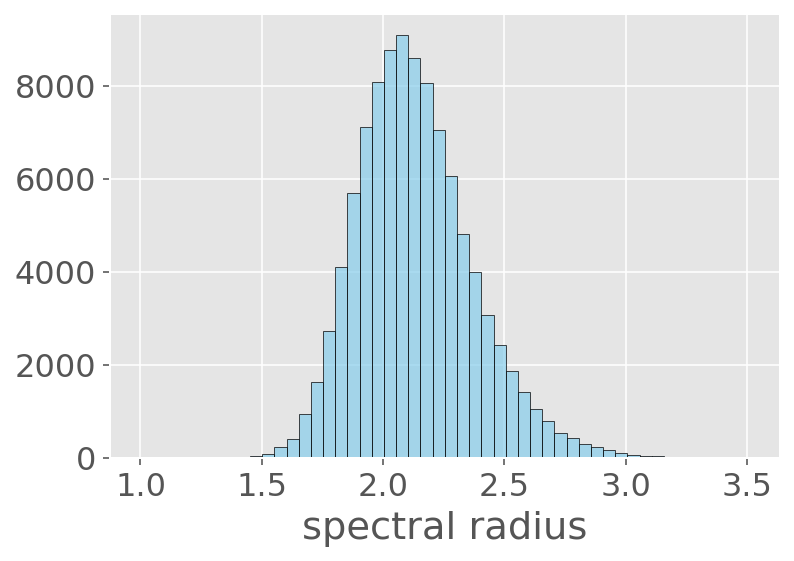} \\
    \small (c) $n=20, p=0.1.$ & \small (d) $n=20, p=0.2$ 
  \end{tabular}
  \caption{Histogram of spectral radii of $B \circ R$ for some values of $(n,p).$ Each has 100,000 samples in 50 bins.}
  \label{fig:spec-rad-histograms}
\end{figure}

{\bf First option for selection of $V$} The first option is to select $V$ to be a constant: $V\equiv v_o$ for some constant $v_o$ that does not depend on $B$ or $R.$ In this case we may wish to select $v_o$ so that $A$ is stable with high probability, or equivalently, $v_o\rho(B \circ R) < 1$ with high probability.  Fig.~\ref{fig:spec-rad-histograms} shows histograms of variates of $\rho(B \circ R)$ for several values of $(n,p).$ The spread of values is considerable so to ensure $A$ is stable with high probability $v_o$ should be chosen to make the expected spectral radius of $A$ considerably less than one.

{\bf Second option for selection of $V$} The second option we consider for choice of $V$ is to set it to make the spectral radius of $A$ equal to a specified constant $r_o,$ where $0< r_o <1.$ In other words, let $V = r_o/\rho(B \circ R).$ Then $Q$ could be taken to be the steady state covariance matrix.  This is the model adopted in \cite{SunTaylorBollt15}.  This choice of $V$ is not well defined on the event $\rho(B \circ R)=0.$ In this paragraph we suggest a way to address that problem.  For $n$ and the average degree $np$ sufficiently large it is very unlikely that $\rho(B \circ R)=0,$ such as in the simulations in \cite{SunTaylorBollt15} where $np=10.$ Moreover, if $\rho(B \circ R)\neq 0$ then $\rho(B \circ R)\geq 1$ (because in that case $(B \circ R)^m$ is a nonzero matrix with integer entries for all $m\geq 1$) and $V \leq r_o$.  However, for small values of $n$ and $np$ there is a significant probability that $\rho(B \circ R)=0.$ That happens if and only if $B \circ R$ is nilpotent which happens, for example, if the directed graph with incidence matrix $B$ has no cycles, including no self loops. Nilpotence of $B \circ R$ can happen in other ways; for example $\left(\begin{array}{cc} 1 & 1 \\ -1 & -1 \end{array}\right)$ is nilpotent.  If $n=10$ and $p=0.2$ we found empirically that $\rho(B \circ R)=0$ with probability about $0.6\%.$ Therefore, if it happens that $\rho(B \circ R)=0$ we need another way to select $V.$ We address that by considering a different approach to scaling that is based on the relationship between the spectral radius of the matrix $A$ and $S$, where $S$ is the average over the $n$ vertices of the variances of the state variables in steady state: $S = \frac 1 n \Tr(Q)$, where $Q$ is the steady state covariance matrix.
\begin{figure}[htb]
  \centering
  \begin{tabular}{c @{\qquad} c }
    \includegraphics[width=.4\linewidth]{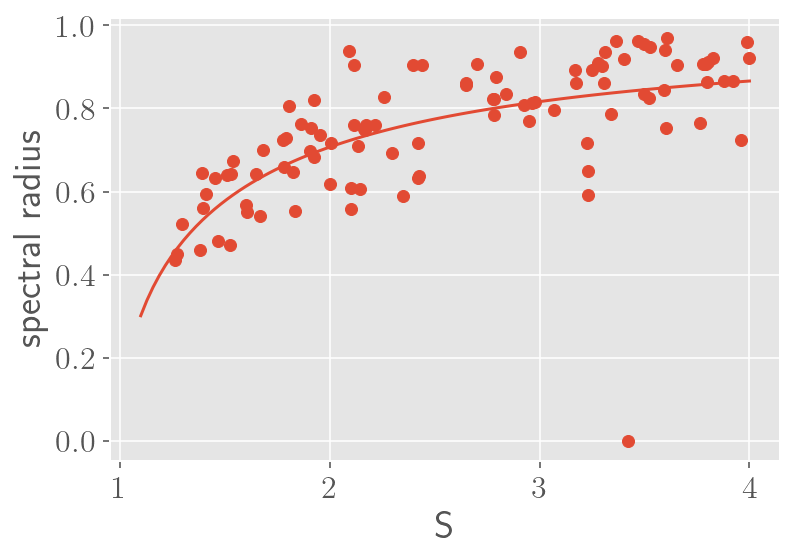} &
    \includegraphics[width=.4\linewidth]{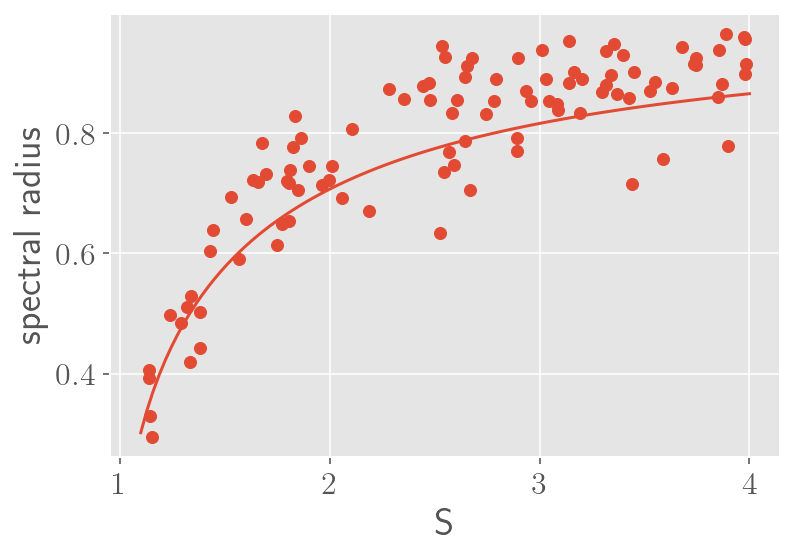} \\
    \small (a) $n=10, p=0.2.$ & \small (b) $n=10, p=0.4$  \\
    \includegraphics[width=.4\linewidth]{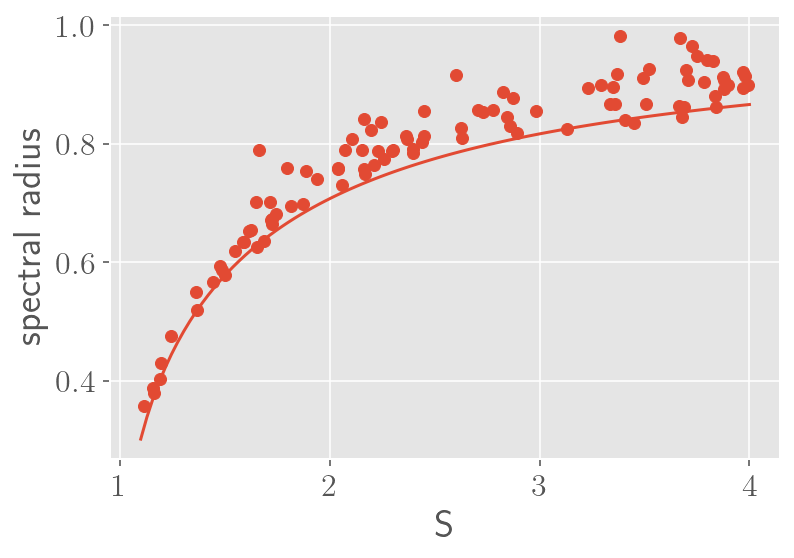} &
    \includegraphics[width=.4\linewidth]{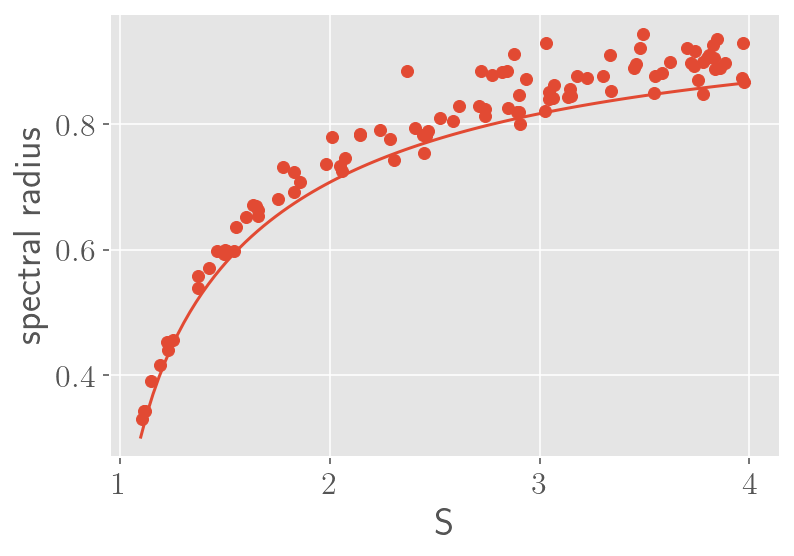} \\
    \small (c) $n=100, p=0.2.$ & \small (d) $n=100, p=0.4$ \\
    \includegraphics[width=.4\linewidth]{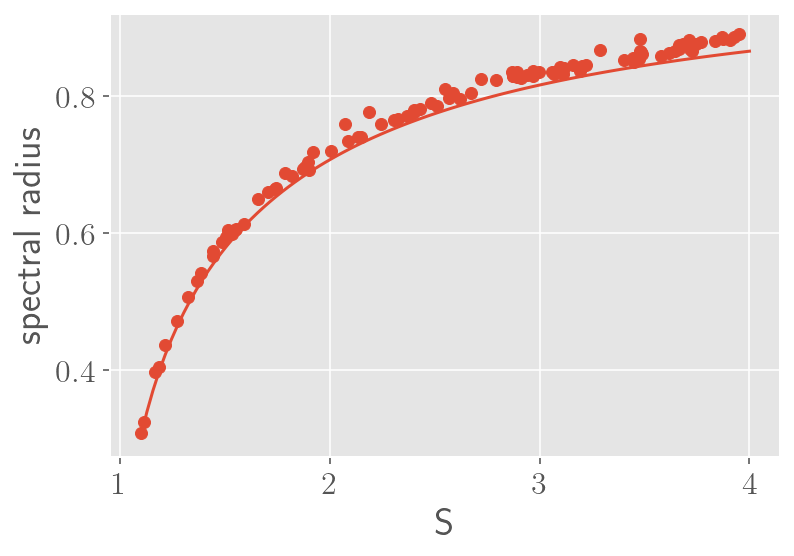} &
    \includegraphics[width=.4\linewidth]{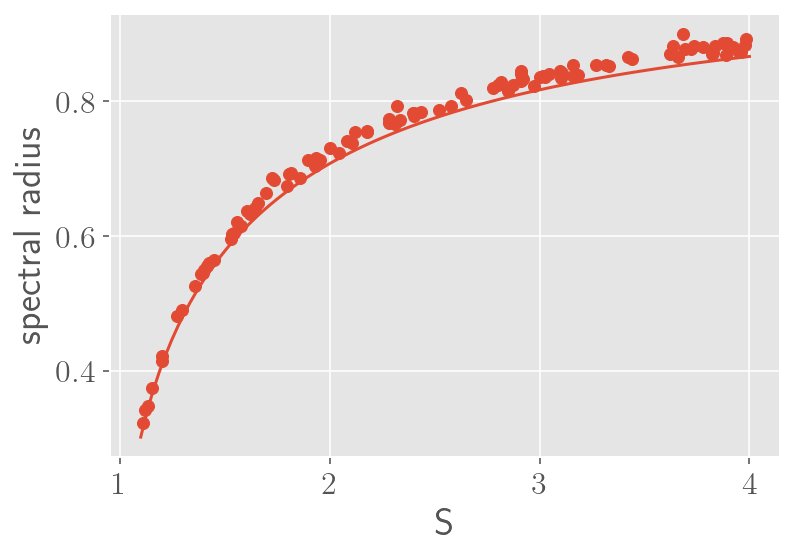} \\
    \small (e) $n=1000, p=0.2.$ & \small (f) $n=1000, p=0.4$ 
  \end{tabular}
  \caption{Spectral radius vs.\ average steady state variance.}
  \label{fig:rho-vs-S_all}
\end{figure}

Fig.~\ref{fig:rho-vs-S_all} shows scatter plots of pairs of $(S,\rho)$ for random instances of the $B \circ R$ matrix.  For each instance of $B \circ R$ a value $S$ was independently generated with the uniform distribution over the interval $[1.1,4]$.  Then the value of $V$ was chosen so that $\frac{1}{n} \Tr(Q) = S.$  With $S$ and $A$ thus determined, the value of $\rho(A)$ was calculated to get the pair $(S,\rho).$  The plots are overlaid with the curve $S = \frac{1}{1-\rho^2}$, or equivalently $\rho=\sqrt{1 - \frac{1}{S}},$ which is the exact relationship for the case of a network with a single vertex (i.e. $n=1$).  It is clear that $S \approx \frac{1}{1-\rho^2}$ for larger networks.  Thus, for a given $r_o$, instead of selecting the scaling constant $V$ so the spectral radius is $r_o$, we could select $V$ so that $S = \frac{1}{1-r_o^2}.$  In our simulations we follow the model of \cite{SunTaylorBollt15} but use the alternative scaling with $S = \frac{1}{1-r_o^2}$ for those variates with $\rho(B \circ R)=0.$  This scaling method still fails if $B\equiv 0$, in which case we replace $B$ by the identity matrix, to be definite.  That occurs with probability $(1-p)^{n^2}$, which is extremely small in most applications.  For example it is about $2\times 10^{-10}$ for $n=10$ and $p=0.2.$

Let $\erPrior$ denote the joint probability distribution of $B$ and $R$.  Since $V$ and hence $A$ is a deterministic function of $B \circ R$, $\erPrior$ induces a distribution on the weighted adjacency matrix $A$.  Note that $\erPrior$ is symmetric in the sense of Remark~\ref{rem:symm}.

\begin{remark}
  Selecting $V$ for a target spectral radius as above has a subtle quirk to be aware of that could be significant in some settings.  Changing $B_{ij}$ to $1-B_{ij}$ for a single ordered vertex pair $(i,j)$ changes $\rho(B \circ R).$  This situation is illustrated in Fig.~\ref{fig:spec-rad-pairs}, which for $n\in\{10,20\}$ and $p\in \{0.2,0.4\},$ shows scatter plots of pairs of spectral radii of $\rho(B \circ R).$
  \begin{figure}[htb]
  \centering
  \begin{tabular}{c @{\qquad} c }
    \includegraphics[width=.4\linewidth]{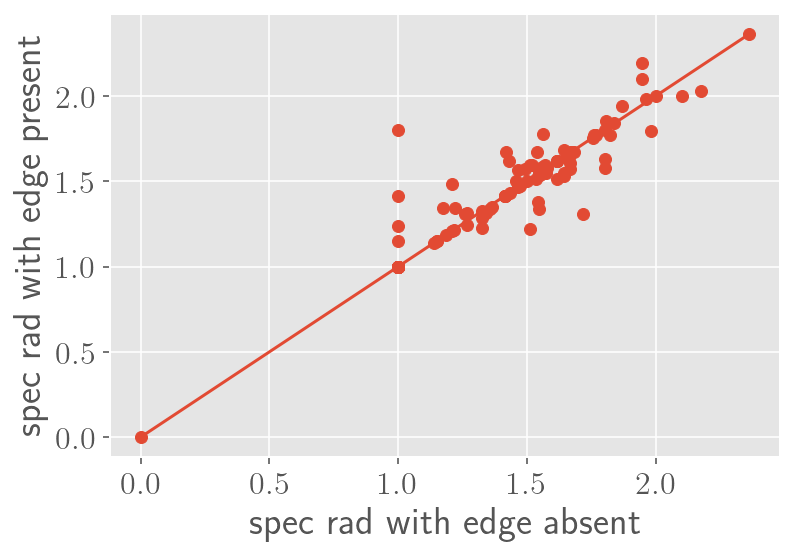} &
    \includegraphics[width=.4\linewidth]{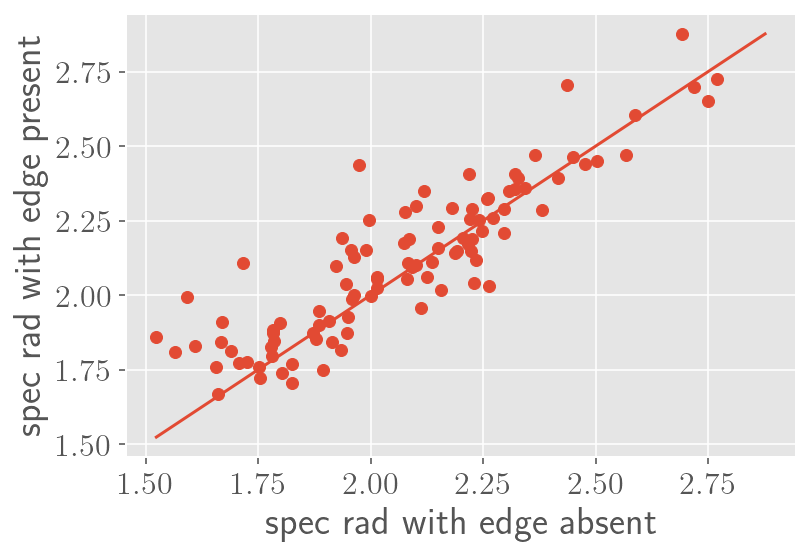} \\
    \small (a) $n=10, p=0.2.$ & \small (b) $n=10, p=0.4$  \\
    \includegraphics[width=.4\linewidth]{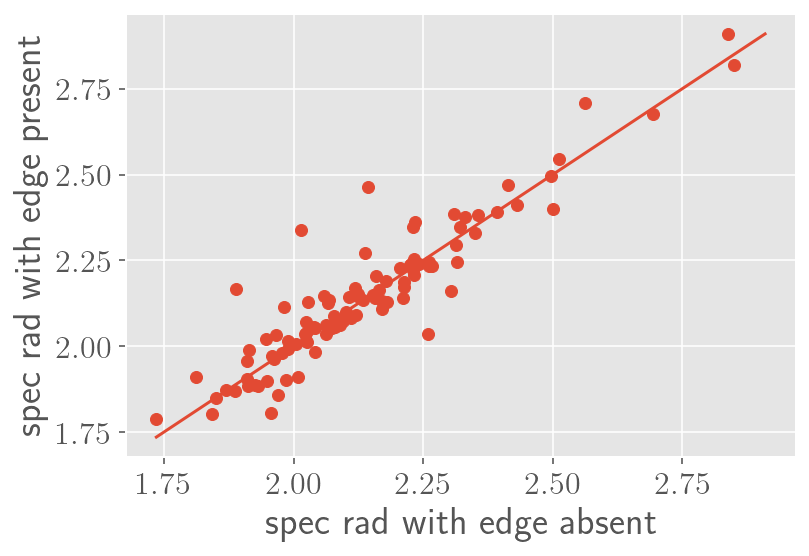} &
    \includegraphics[width=.4\linewidth]{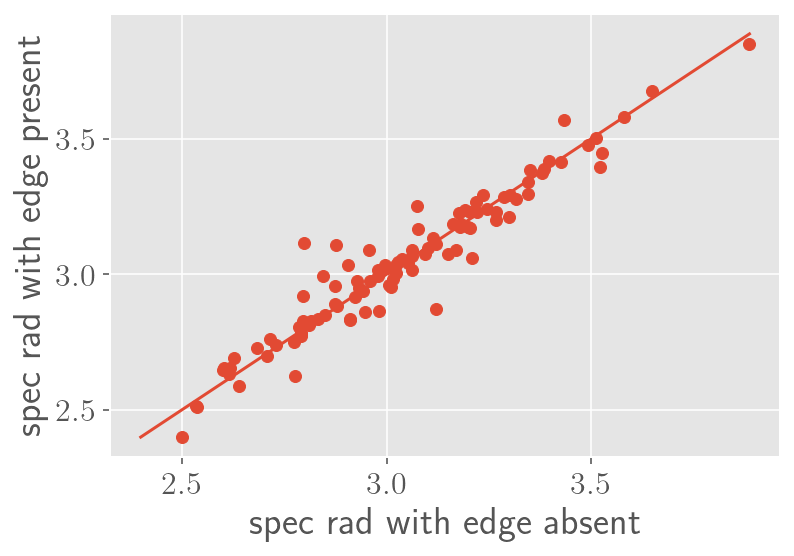} \\
    \small (c) $n=20, p=0.2.$ & \small (d) $n=20, p=0.4$ 
  \end{tabular}
  \caption{Scatterplots of spectral radius pairs for some values of $(n,p).$}
  \label{fig:spec-rad-pairs}
\end{figure}

  Each pair corresponds to an instance of $\rho(B \circ R)$ with the status of a fixed pair of vertices set to either edge present or edge absent.  The plots show that the spectral radius can change significantly with the addition or deletion of a single edge.  This points to a quirk of Definition~\ref{def:er} for the causal inference application.  Namely, since the network is scaled by a factor $V$ to have spectral radius precisely $r_o$, the scaling $V$ of every edge in the network changes (usually slightly) if a single edge is changed from present to absent or vice versa.  This effect is small and shouldn't be of concern unless the variance $\nu^2$ of the observation noise is very small or zero, $r_o$ is close to one, or $T$ is large.  We view this quirk in the model as the price for having a random model for the network with a specified spectral radius.  The importance of the spectral radius on inference problems has been recognized in previous works, such as \cite{SimchowitzManiaTu18} and \cite{JedraProutiere23}, which address the problem of estimating $A$ (rather than the support of $A$ as in this paper).  For example a minimax lower bound on sample complexity in \cite{SimchowitzManiaTu18} is for $A$ ranging over matrices of the form $\rho$ times a real unitary matrix for a constant $\rho \in (0,1).$
\end{remark}

\section{Edge detection with possible side information}
\label{sec:genie_bounds}
Motivated by Proposition~\ref{prop:network-edge} we consider binary hypothesis testing problems concerning the possible existence of an edge for a given directed pair of vertices.  We shall assume that the parameters $p, \sigma^2, \nu^2$ and $v_o$ or $r_o$ are known to the estimators and omit them from the notation.  We also assume the estimators know which option is used for the selection of the scaling factor $V$.  Also, to be definite, if the first option, $V=v_o$ is used, we assume the initial state is $X(0)=0$ and if the second option is used we assume $X(0)$ has the equilibrium distribution.  Of course the lower bounds to error probability we derive hold also for estimators without access to such knowledge.

\subsection{Optimal detection with no side information}
The distribution of the observed vector $Y=(Y(0), Y(1),\ldots , Y(T))$ is conditionally mean zero and Gaussian given the weighted adjacency matrix $A$ or given the pair $(B,R).$ Since $A$ is random we thus have a Gaussian mixture model.  The model is symmetric up to permutation of the vertices so for simplicity we focus on detecting the existence of a directed edge from vertex 2 to vertex 1.  Let $H_0$ denote the hypothesis that such edge does not exist and $H_1$ denote the hypothesis that such edge does exist.  By the Neyman-Pearson lemma, the Pareto front of decision rules for deciding $H_1$ vs. $H_0$ is given by the likelihood ratio test, based on the likelihood ratio $L(y) = f_1(y)/f_0(y)$ where for $b\in \{0,1\}$, $f_b$ is the density of $Y$ given $H_b$ is the true hypothesis.  Borrowing terminology from game theory, let $B_{-21}$ be all the entries of $B$ except $B_{21}$:
\[B_{-21}= (B_{i'j'}: i',j' \in [n], ~(i',j')\neq (2,1)).\]
Let $\Sigma_{B,R}$ denote the $n(T+1)\times n(T+1)$ covariance matrix of $Y$ given $(B,R).$ Each $f_b$ is a mixture of a large number of Gaussian densities:
\begin{align}
  \label{eq.likihood_ratio_sum}
  f_b(y) = \sum_{B,R:B_{21}=b}f(y|B,R) \erPrior(B_{-21},R)
\end{align}
where $f(y|B,R)$ is the density of the ${\cal N}(0_{1\times n(T+1)}, \Sigma_{B,R})$ distribution and $\erPrior(B_{-21},R)$ is the marginal probability mass function of $(B_{-21},R)$ under $\erPrior$.  The density $f(y|B,R)$ depends on $B,R$ only through the Hadamard product $B \circ R$.  Since there are three possibilities for each entry of $B \circ R$ the total number of distinct possible values of $B \circ R$ is $3^{n^2}$ so that numerically computing the sum in \eqref{eq.likihood_ratio_sum} is computationally prohibitive unless $n$ is very small.

\subsection{All but one side information}
\label{sec:all-but-one}
In order to reduce the complexity we endeavor to compute an upper bound on the performance by assuming the decision maker has access to side information provided by a genie in making the decision.  Since the decision maker can ignore the genie, having access to the genie can't decrease the possible performance.  Thus, if the decision rule makes optimal use of the side information provided by the genie then the corresponding ROC for the decision rule is an upper bound to the ROC for any rule with access only to the observation $Y.$

Perhaps the most simple genie to use is one that provides $(B_{-21}, R).$ We term this the {\em all but one} side information because it reveals whether all edges exist except the one edge being detected.  With this side information, the conditional distribution of $Y$ given a hypothesis $H_b$ for $b\in \{0,1\}$ and $(B_{-21}, R)$ is a mean zero multivariate Gaussian random vector so that the conditional likelihood ratio is readily computed.  Therefore a random variate (or sample) of the likelihood ratio can be calculated by evaluating the likelihood ratio function at a random variate of $Y$ generated under either $H_0$ or $H_1.$

Note that under the modeling option with $V=r_o/\rho(B \circ R)$ the genie does {\em not} reveal the value of the scaling factor $V$.  If it did, $V$ and $(B_{-21}, R)$ would determine $B_{21}$ in the highly likely case that $\rho(B \circ R)$ changes if $B_{21}$ is replaced by $1-B_{21}.$ The genie information would enable perfect detection with no need to use the observations $Y!$

\section{Computing ROC bounds from likelihood ratio samples}
\label{sec:computing_bnds}
The (optimal) receiver operating characteristic (ROC) curve for a detection problem is a graph of the optimal correct edge detection rate as a function of the false positive rate (aka probability of detection as a function of the probability of false alarm).  A method to accurately estimate an ROC curve for a BHT based on samples of the likelihood ratio is described in \cite{HajekKang25}.  If the likelihood ratios are based on actual observations and also side-information provided by a genie then the resulting ROC is an estimate of an upper bound on the true ROC that is increasingly accurate as the number of likelihood ratio samples increases.

\subsection{Plugin generation of likelihood ratio variates with all but one side information}
Consider the all but one genie information described in Section~\ref{sec:all-but-one}.  The genie provides $R,B_{-21}.$  Independent samples of likelihood ratio variates can be computed under either hypothesis $H_b:B_{2,1}=b,$ for $b\in\{0,1\}.$  Those samples can then be used in the maximum likelihood estimator from \cite{HajekKang25}.  Algorithm~\ref{alg:LR_sample} shows a method for generating a likelihood ratio sample that we call the plugin method.  First a random weighted adjacency matrix $A$ is generated, then the pdf of random vector $Y$ is generated with $B_{21}$ set to 0 or 1, and then the ratio of the densities is evaluated at a randomly generated sample of $Y.$

\begin{algorithm}
  \caption{Plugin generation of a likelihood ratio variate}
  \label{alg:LR_sample}
  \begin{algorithmic}
    \REQUIRE{$n\geq 2,p\in [0,1],(v_o,Q) \mbox{ or } r_o,b\in\{0,1\}, \sigma^2> 0, \nu^2 > 0$}\\
    \STATE Generate a variate of $B,R$ given $n,p$
    \FOR{$k \in \{0,1\}$}
    \STATE $B_{1,2}\gets k$
    \STATE $\Sigma_k \gets \cov(Y)$ using $B,R,(v_o,Q) \mbox{ or } r_o,\sigma^2,\nu^2$\\
    \ENDFOR
    \STATE $y \gets$ a $N(0, \Sigma_b)$ variate
    \FOR{$k \in \{0,1\}$}
    \STATE $f_k(y) \gets$  $N(0, \Sigma_k)$ pdf evaluated at y
    \ENDFOR
    \RETURN $R = \frac{f_1(y)} {f_0(y)}$
  \end{algorithmic}
\end{algorithm} 

The method of Algorithm~\ref{alg:LR_sample} can be applied for either modeling option for the choice of $V$ described in Section~\ref{sec:network-topology}.  The method requires computation of joint Gaussian densities in $(n+1)T$ dimensions which involves the inverses of $(n+1)T \times (n+1)T$ covariance matrices.  We found this to be numerically stable for $nT$ at least as large as 200, thanks to the log-density techniques used in standard numerical packages such as scipy.stats in python.

\subsection{Innovation generation of likelihood ratio samples with all but one side information and no observation noise}
Calculation of the pdfs in Algorithm~\ref{alg:LR_sample} requires inversion of the $(n+1)T \times (n+1)T$ covariance matrices.  That can be avoided in the case of no observation noise (i.e. $\nu^2=0$) and for the first modeling option for the choice of $V$ -- namely $V\equiv v_o$ for some constant $v_o.$  The idea is to focus on the innovation, or new information, at each time step and exploit factorization of the probability density of the observations.  Suppose there is no observation noise so that $X = (X(t): 0 \leq t \leq T)$ is observed.  We focus on the single directed pair of vertices $(2,1)$ and suppose a genie provides the value $(B_{-21}, R).$  For $b\in\{0,1\}$, under the hypotheses $H_b :B_{21} = b$ the likelihood of observing $X=x$ given $(B_{-21}, R)$ is
\begin{align}
  \label{eq:factorization_with_no_noise}
  f_b(x | B_{-21}, R) & = \prod_{t = 1}^T \left[f(x_{-1}(t)\mid x(t - 1))\cdot f_b(x_1(t) \mid x(t - 1))\right]
\end{align}
where $x_{-1}(t)=(x_2(t), \ldots , x_n(t))$ and the conditional density functions $f_b$'s are given by the conditional distributions
\[\law_b(x_1(t)\mid x(t - 1)) = \mathcal N\left(\left(v_o\sum_{k \neq 2}B_{k1}R_{k1}x_k(t - 1)\right) + v_obR_{21}x_2(t - 1), \sigmab^2\right).\]
The likelihood ratio is
\begin{align}
  \lr & \defeq\frac{f_1(x| B_{-21}, R)}{f_0(x| B_{-21}, R)}\nonumber \\
      & = \exp\left[\frac 1{2\sigmab^2}\sum_{t = 1}^{T}\left(\tilde x_1(t)^2 - \left(\tilde x_1(t) - v_oR_{21}x_2(t - 1)\right)^2\right) \right] \nonumber\\
      & = \exp\left[\frac 1{\sigmab^2}\sum_{t = 1}^T v_oR_{21}x_2(t - 1)\left(\tilde x_1(t) - \frac 12v_oR_{21}x_2(t - 1)\right) \right], \label{eq:LR_sample_no_obs_noise}
\end{align}
where
\begin{align}
  \tilde x_1(t) & = x_1(t) - v_o\sum_{k\neq 2}B_{k1}R_{k1}x_k(t - 1). \label{eq:xtilde}
\end{align}
Algorithm~\ref{alg:LR_sample_innovation} summarizes how to generate a variate of the likelihood ratio under hypothesis $H_b$ for $b\in \{0,1\}$ for all but one side information.
\begin{algorithm}
  \caption{Innovation generation of a likelihood ratio variate}
  \label{alg:LR_sample_innovation}
  \begin{algorithmic}
    \REQUIRE{$n\geq 2,p\in [0,1], v_o, Q,  b\in \{0,1\}, \sigma^2> 0$} \\
    \STATE Generate a variate of $B,R$ given $n,p$ \\
    \STATE $B_{21} \gets b$ \\
    \STATE $A \gets v_oB \circ R$ \\
    \STATE Generate $X=(X(t): 0 \leq t \leq T)$ using $Q,A,\sigma^2$ according to \eqref{eq:state_evolution} \\
    \STATE Calculate $\lr$ using \eqref{eq:LR_sample_no_obs_noise} and \eqref{eq:xtilde} \\
    \COMMENT {Previous two steps can be done simultaneously -- one term in sum in \eqref{eq:LR_sample_no_obs_noise} can be added just after generating $X(t)$}
    \RETURN $\lr$
  \end{algorithmic}
\end{algorithm}

\section{Numerical results}
\label{sec:num}

Numerical results from simulations are presented in this section that compare the performance of two algorithms to the upper bound on ROC provided by all-but-one side information.

\subsection{Algorithms}
Let
\[\Phi(0) =
  \begin{pmatrix}
    Y(0)\\Y(1)\\\vdots\\Y(T-1)
  \end{pmatrix},\quad \Phi(1) =
  \begin{pmatrix}
    Y(1)\\Y(2)\\\vdots\\Y(T)
  \end{pmatrix}.
\]

\subsubsection{lasso}
The lasso algorithm solves the optimization problem
\[\minimize_{A_j}\frac 1{2T}\|\Phi_j(1)-\Phi(0)A_j\|_2^2 + \lambda\|A_j\|_1,\]
where $A_j$ and $\Phi_j(1)$ are the $j$th columns of $A$ and $\Phi(1)$, respectively, and $\lambda \ge 0$ is the regularization parameter.  If $\Phi(0)\T\Phi(0)$ is invertible, the minimizer $\hat A_j^{\mathsf{lasso}}$ is unique.  Write $\hat A_j^{\mathsf{lasso}} = \ind{\hat A_{ij}^{\mathsf{lasso}}}{i\in[n]}$.  Then the estimated support matrix is $\Best^{\mathsf{lasso}}$ defined by $\Best_{ij}^{\mathsf{lasso}} = \indicator{\{\hat A_{ij}^{\mathsf{lasso}}\neq 0\}}$.  We implement lasso using \textit{scikit-learn} \cite{PedregosaVaroquauxGramfort11}.

\subsubsection{oCSE}
oCSE was proposed in \cite{SunTaylorBollt15}.  For each target vertex $j$, its parent set is discovered greedily one at a time by finding the vertex whose column in $\Phi(0)$ together with the other chosen columns fits $\Phi_j(1)$ the best in the least squares sense.  This discovery stage terminates when the improvement in the residual fails a permutation test with some threshold $\theta.$

\subsection{Numerical performance}
\begin{figure}[htb]
  \centering
 \begin{tabular}{c c }
    \includegraphics[width=.45\linewidth]{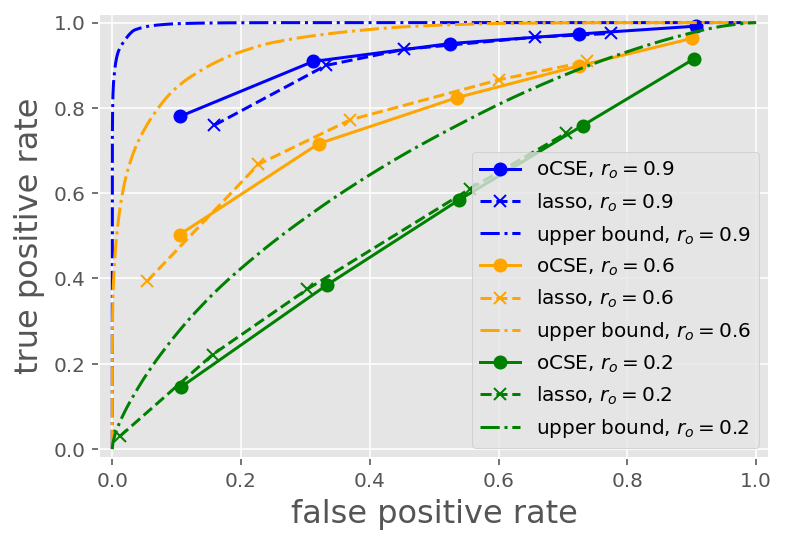} &
    \includegraphics[width=.45\linewidth]{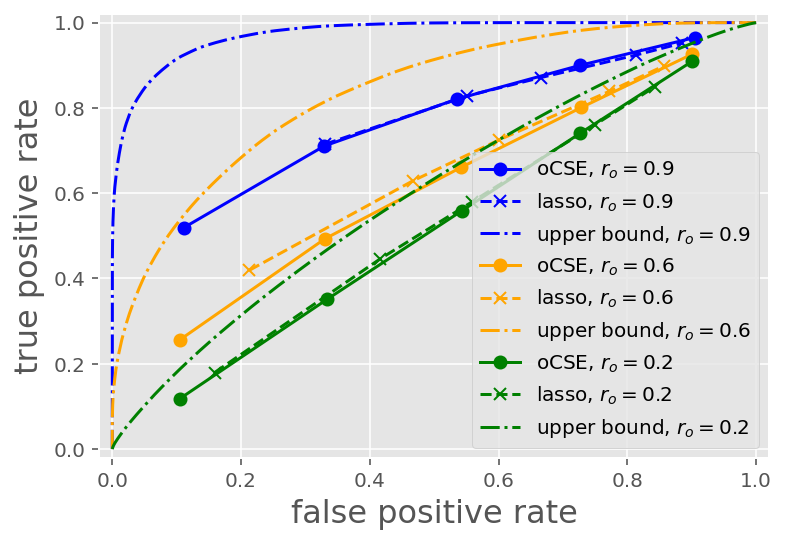} \\
    \small (a) No observation noise ($\nu^2 = 0).$ & \small (b) Observation noise with $\nu^2=1.$
  \end{tabular}
  \caption{ROC curves of oCSE and lasso and the ML ROC all-but-one upper bound for $n = 10$, $p = 0.2$, $T = 20$, and scaling to spectral radius averaged over 500 simulations.}
  \label{fig:ocse-lasso-MLROC-stationary}
\end{figure}

Comparisons of the receiver operating characteristic (ROC) curves of lasso and oCSE by varying the parameters $\lambda$ and $\theta$ and the upper bounds on the ROC curve resulted from Proposition~\ref{prop:network-edge} for the ternary \er\ model with the stationary initial condition on $X(0)$ are shown in Fig.~\ref{fig:ocse-lasso-MLROC-stationary}.  Subplot~\ref{fig:ocse-lasso-MLROC-stationary}(a) is for no observation noise (i.e. $\nu^2 =0)$ and Subplot~\ref{fig:ocse-lasso-MLROC-stationary}(b) is for observation noise with $\nu^2 =1.$  The upper bound curves for these figures were produced using the plugin method, so the network size ($n=2$) and time duration ($T=20$) is near the limit of the method.  We could also use the innovation method for the first figure and get the same result.  The effect of the observation noise is to shift the ROC curves downward as expected.  There is a considerable gap between the upper bound and the algorithm performance.  It could be due to the use of the genie in the upper bound or to suboptimality in the algorithms.

\begin{figure}[htb]
  \centering
  \includegraphics[width=\figWidth]{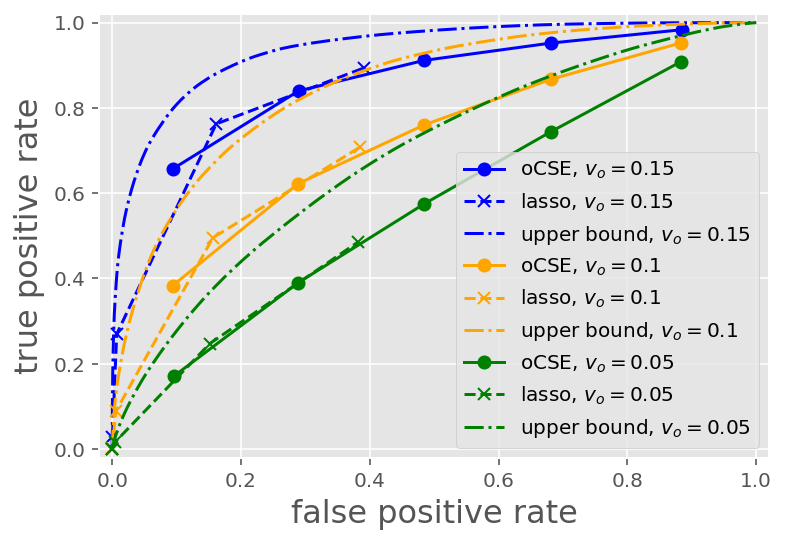}
  \caption{ROC curves of oCSE and lasso and the ML ROC all-but-one upper bound for $n = 20$, $p = 0.1$, $T = 200$, and $\sigmat^2 = 0$ and constant scale factors, averaged over 500 simulations.}
  \label{fig:ocse-lasso-MLROC-stationary_False-num_genes_20-num_times_200-sims_100-prob_conn_0.2-obs_noise_0.png}
\end{figure}

A similar comparison is shown in Fig.~\ref{fig:ocse-lasso-MLROC-stationary_False-num_genes_20-num_times_200-sims_100-prob_conn_0.2-obs_noise_0.png} for the zero initial state version of the model.  Considerably larger networks can be simulated for this version of the model because we can use the innovation method for generating likelihood ratio samples.  For this larger network we also see a considerable gap between the upper bound and algorithm performance.

\section{Upper bounds on ROC for causal inference using information theoretic measures}
\label{sec:bounds_casual-recov-info-measures}
In the previous section the performance of two algorithms was compared to an upper bound on the ROC based on the all but one side information and ROC curve generated based on likelihood ratio samples.  Much work on the performance analysis of estimation and detection bounds is based on information theoretic measures between two probability distributions.  Use of such measures has many advantages, especially in asymptotic analysis, but there is a gap between the true ROC curve and the ROC curve implied by such estimators.  To get an idea of what that gap would be in the context of the causal inference problem of this paper, in this section we compare the true ROC (as estimated from likelihood ratio samples) to ROC bounds generated using Kullback--Leibler divergence or Bhattacharyya coefficients.  See Appendix~\ref{app:info_bnds} for notation and a brief summary of the use of such bounds in a simpler but related context -- namely, the detection of a random signal in noise.

\subsection{Plugin generation of information theoretic measures with all but one side information}
Samples of the conditional $BC$ or $d_{KL}$ given all-but-one side information can be generated by a modification of Algorithm 1.  Once $\Sigma_0$ and $\Sigma_1$ are computed the algorithm returns the BC or $d_{KL}$ for the two distributions ${\cal N}(0,\Sigma_b)$ and ${\cal N}(0,\Sigma_{1-b})$ by plugging into the following formulas:
\begin{align}
  BC^2({\cal N}(0,\Sigma_1)\| {\cal N}(0,\Sigma_0)) &= \frac{\sqrt{\det \Sigma_1\det\Sigma_2}}{\det\frac{\Sigma_1 + \Sigma_2}2} \\
  D_{KL}({\cal N}(0,\Sigma_1)\| {\cal N}(0,\Sigma_0)) &= \frac 1 2 \left[ tr\left( \Sigma_1^{-1} \Sigma_0\right) - n + \ln \frac{\det \Sigma_1 }{\det \Sigma_0}\right]
\end{align}

\subsection{Innovation generation of KL distance samples with all but one side information and no observation noise}
In the case of no observation noise and mean zero Gaussian initial state $X(0)$ the distances $D_{KL}(P_0\|P_1)$ and $D_{KL}(P_1\|P_0)$ can be efficiently computed given $v_o$ using the following formula:
\begin{align}
  D_{KL}(P_b \| P_{1-b}) = \sum_{t=0}^{T-1}  \frac {v_o^2} 2 E_b[X_2^2(t)]   \label{eq:KL_formula}
\end{align}
where $E_b[X_2^2(t)] = \cov (X(t))_{2,2}$ and $\cov (X(t))$ can be computed recursively by $\cov (X(t+1)) = A_b\cov (X(t))A_b^T + I.$  Formula \eqref{eq:KL_formula} follows from the factorization \eqref{eq:factorization_with_no_noise} and is a special case of computation of KL distances between two Markov sequences using the chain rule of KL divergence \cite{CoverThomas06}.  Equation \eqref{eq:KL_formula} is a special case of an expression derived in \cite{JedraProutiere19}.

\subsection{Numerical comparison}
\begin{figure}[htb]
  \centering
  \begin{tabular}{c c }
    \includegraphics[width=.45\linewidth]{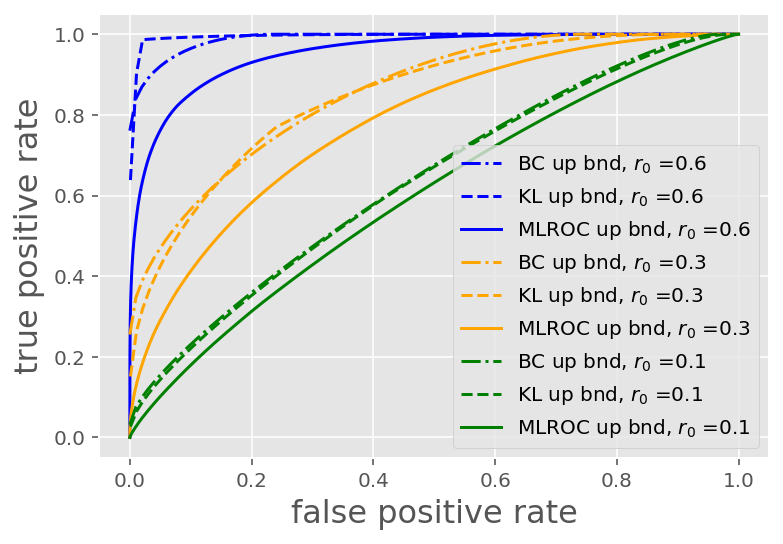} &
    \includegraphics[width=.45\linewidth]{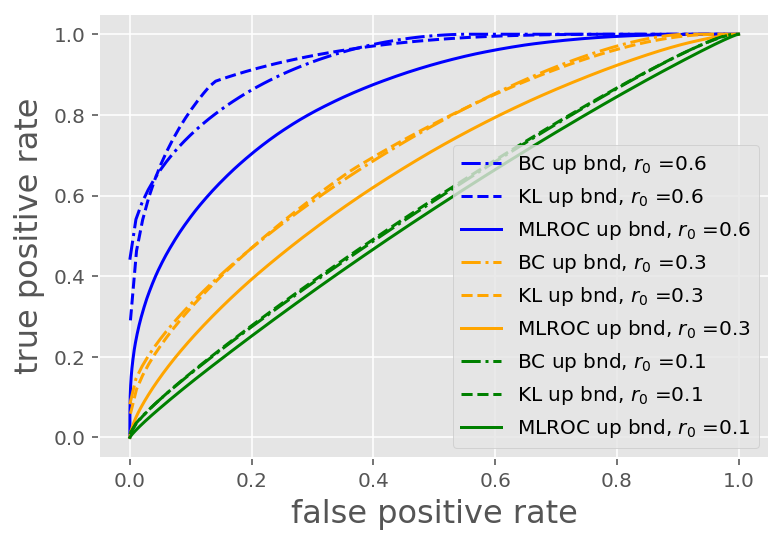} \\
    \small (a) No observation noise ($\nu^2 = 0).$ & \small (b) Observation noise with $\nu^2=1.$
  \end{tabular}
\caption{ROC curves of ML ROC, BC bound, and KL bound for all-but-one upper side information for $n = 10$, $p = 0.2$, $T = 20$, and scaling to spectral radius averaged over 1000 simulations.}
\label{bnd_compare-stationary}
\end{figure}

Fig.~\ref{bnd_compare-stationary} shows a comparison of the ROC curves of ML ROC, BC bound, and KL bound ROC curves for all-but-one side information for $n = 10$, $p = 0.2$, $T = 20$, and scaling to spectral radius averaged over 1000 simulations.

\begin{figure}[htb]
  \centering
  \includegraphics[width=\figWidth]{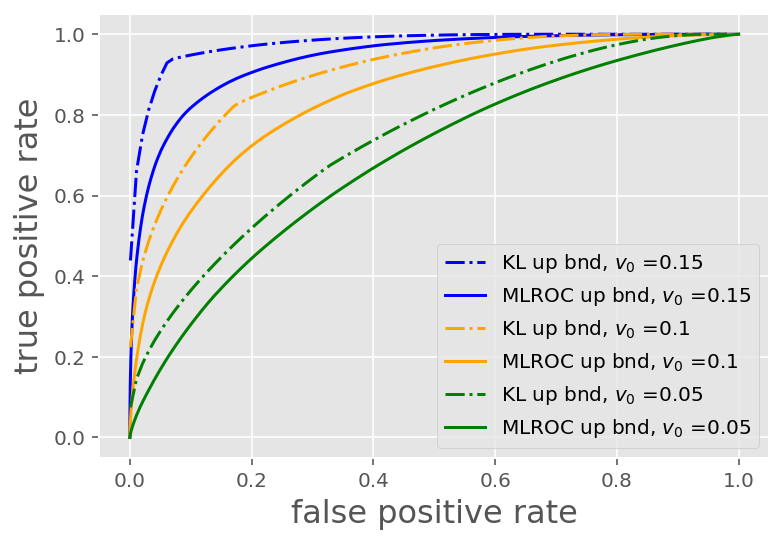}
  \caption{ROC curves of ML ROC and KL bound for all-but-one upper side information for $n = 20$, $p = 0.1$, $T = 200$, $\sigmat^2 = 0$ and constant scale factors, averaged over 1000 simulations.}
  \label{bnd_compare-transient}
\end{figure}

A similar comparison is shown in Fig.~\ref{bnd_compare-transient} for the zero initial state version of the model for a larger network with $n=20$ and $T=200.$  The ML ROC curve and KL bound are shown; they were computed using the innovation method which doesn't work for the BC bound.

Overall, examination of Figs.~\ref{bnd_compare-stationary} and \ref{bnd_compare-transient} indicates that the bounds implied by the information theoretic measures are close to the actual ROCs (all for the case of all but one side information in the causal inference problem).  See the appendix for additional discussion of the difference between the true ROC and the ROCs implied by the information theoretic measures.

\section{Discussion}
\label{sec:discussion}

As mentioned in Section~\ref{sec:num} there is a significant gap between the ROC upper bound and the ROCs of the algorithms.  On one hand, the source of the gap could be the large amount of side information that the genie provides.  Therefore it may be of interest to explore methods that do not involve side information or provide less side information.  A challenge in defining slightly less side information (i.e., a weaker genie with more than one alternative hypothesis) is to identify multiple plausible alternative hypotheses in order to stress the decision rule.  On the other hand, the source of the gap could be suboptimality of the algorithms.  There are many existing algorithms and undoubtedly many more can be devised based on machine learning.

This paper has focused on the small signal-to-noise ratio regime in the sense that the ROC curves we have examined do not get very close to the ideal performance point $(1,1)$.  Another avenue for further research is to focus on the high signal-to-noise ratio regime.  Perhaps the all-but-one side information offers a tighter bound in such cases.

We have carefully defined the model of this paper so it could be used as a benchmark, but at least in the domain of biological networks there are much more complex network models (see \cite{KangHajekHanzawa20} for example).  Just for one example, the additive nature of the Gauss-Markov model used in this paper treats nodes somewhat like logical OR gates whereas biological networks can have behavior more similar to AND gates.

\appendices
\section{Information theoretic upper bounds on ROC curves}
\label{app:info_bnds}

Various information measures between two probability distributions can be used to bound error probabilities for BHT problems.  To give an indication of how tight the bounds are we look at numerical examples involving Bhattacharyya coefficients and Kullback--Leibler (KL) divergences.  Consider a BHT problem such that observation $Y \in {\cal Y}$ has probability distribution $P$ under $H_1$ and distribution $Q$ under $H_0.$  For any decision rule $\delta\colon \mathcal Y \to \{0, 1\}$, the probability of detection and probability of false alarm are given by $p_{det} = E_P[\delta(Y)]$ and $p_{fa} = E_Q[\delta(Y)]$.  The KL divergence is defined by $D_{KL}(P\|Q) = \int\left( \log\frac{dP}{dQ}\right)dP$ and the squared Hellinger distance is defined by $H^2(P,Q) = \int \left(\sqrt{dP}-\sqrt{dQ}\right)^2 = 2(1-BC(P,Q))$ where $BC(P,Q)$ is the Bhattacharyya coefficient: $BC(P,Q) = \int \sqrt{dPdQ}$.

Given a convex function $f$ on $(0,\infty)$ with $f(1)=0$ the $f$-divergence between $P$ and $Q$ is given by $D_f(P\|Q) = \int f\left( \frac{dP}{dQ} \right) dQ.$  See \cite{Csiszar67} and \cite{Polyanskiy_Wu_2025} for background.  The KL divergence is an $f$-divergence for $f(u)=u\log u$ and $H^2$ is an $f$ divergence for $f(u) =\left(\sqrt{u}-1\right)^2.$

The $f$-divergences satisfy the data processing inequality so in particular $d_f(p_{det} \|p_{fa}) \leq D_f(P\| Q ),$ where $d_f(p \|q)$ is the $f$ divergence between the Bernoulli distributions with parameters $p$ and $q.$  Since $H^2 = 2(1-BC),$ the $BC$ also satisfies the data processing theorem with the inequality reversed.  Therefore: $bc(p_{det},p_{fa}) \geq BC(P,Q)$ where $bc(p,q) =\sqrt{pq} + \sqrt{(1-p)(1-q)}.$  This observation yields the following tight upper bounds on ROC curves implied by $BC(P\|Q)$ or $D_f(P\|Q),$ respectively:
\begin{align}
  p_{det}& \leq \max \left\{p \in [p_{fa},1]: bc(p, p_{fa} )\geq BC(P, Q) \right\}\label{eq:BC_ROC_bound}\\
  p_{det}& \leq \max \left\{p \in [p_{fa},1]: d_f(p \| p_{fa} )\leq D_f(P\| Q)\mbox{ and }d_f(p_{fa} \| p ) \leq D_f(Q\| P) \right\}.\label{eq:KL_ROC_bound}
\end{align}
If $P\sim {\cal N}(\mu_1,1)$ and $Q\sim {\cal N}(\mu_0,1)$, then $BC(P,Q) = e^{-(\triangle \mu)^2/8}$ and $d_{KL}(P,Q) = (\triangle\mu)^2/2$ where $\triangle \mu = \mu_1 - \mu_0.$

If $(X,Y)$ is a pair of random variables such that the marginals $P_X$ and $Q_X$ are the same, then
\begin{align}
  D_f(P_{X,Y} \| Q_{X,Y}) &= \int D_f(P_{Y|X=x} \| Q_{Y|X=x}) P_X(dx)   \label{eq:Df_tower_rule} \\
  BC(P_{X,Y} \| Q_{X,Y}) &= \int BC(P_{Y|X=x} \| Q_{Y|X=x}) P_X(dx).   \label{eq:BC_tower_rule}
\end{align}

We examine the tightness of the upper bounds \eqref{eq:BC_ROC_bound} and \eqref{eq:KL_ROC_bound} for the simple case of detection of a known random signal in noise.  Suppose $X$ and $W$ are mutually independent random vectors in $\reals^n$ such that $W\sim \mathcal N(0,I).$  Consider the binary hypothesis testing problem with observation $(X,Y)$: $H_0: Y=W$ vs. $H_1: Y= X+W.$  The likelihood ratio is
\begin{align*}
  \lr = \exp\left(\left\langle X, Y-\frac12 X\right\rangle\right).
\end{align*}
The Bhattacharyya coefficient and KL divergences for the conditional distribution of $Y$ given $X=x$ are as follows:
\begin{align}
  \label{eq:conditional_measures}
  BC(P_{Y|X=x},Q_{Y|X=x}) = e^{-\|x\|^2/8}~~~  D_{KL}(P_{Y|X=x}\|Q_{Y|X=x}) = \frac {\|x\|^2} 2.
\end{align}
As long as we can generate independent samples of $X$ and $W$ we can generate samples of the likelihood ratio under either hypothesis and then estimate the ROC using \cite{HajekKang25}.  Also, with samples of $X$ we can use \eqref{eq:conditional_measures} to accurately estimate the integrals in \eqref{eq:Df_tower_rule} and \eqref{eq:BC_tower_rule} by replacing the integration with respect to $P_X$ by integration with respect to the empirical distribution of $X.$  We can therefore accurately estimate the Bhattacharyya coefficient and KL divergences and compute the corresponding upper bounds on the ROC according to \eqref{eq:BC_ROC_bound} and \eqref{eq:KL_ROC_bound}.  This is done for two examples.  Both the ROC curve and the bounds depend on the distribution of $X$ only through the distribution of $\|X\|.$

\begin{exam}[Constant norm signal -- binormal detection]
  Suppose the norm of $X$ is deterministic: $\|X\|=\mu$ with probability one for some $\mu > 0.$  The true ROC is given parametrically by $\gamma \mapsto (1-\Phi(\gamma), 1-\Phi(\gamma-\mu))$, where $\Phi$ is the CDF of the standard Gaussian distribution.  Also, $BC = e^{-\mu^2/8}$ and $D_{KL}(P_1\|P_0) = D_{KL}(P_0\|P_1) =\frac{\mu^2}{2}.$  Fig.~\ref{fig:binormROC} shows the ROC together with the upper bounds based on the Bhattacharyya coefficient and the KL divergence.
  \begin{figure}[thbp]
    \centering
    \includegraphics[scale=0.7]{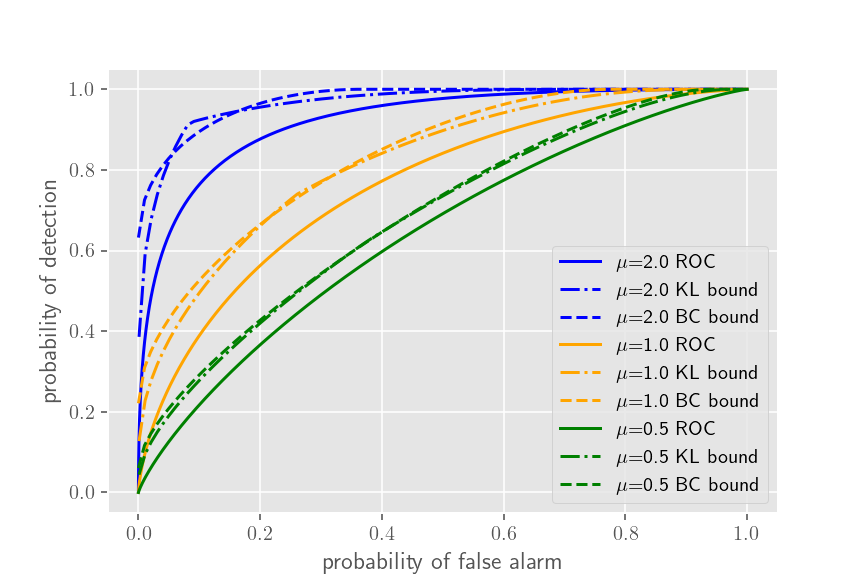}
    \caption{ROC and bounds for $\|X\|\equiv \mu.$}
    \label{fig:binormROC}
  \end{figure}
  The upper bounds are close to each other and there is a noticeable gap between them and the true optimal ROC.
\end{exam}

\begin{exam}[Markov signal]
  Suppose $X$ is a length $n$ segment from a symmetric time-homogeneous Markov process on states $\{0, A\}$ for some constant $A>0$, with crossover transition probability $p$ and initial distribution $(0.5,0.5).$  Fig.~\ref{fig:markovROC} shows the ROC together with the upper bounds based on the Bhattacharyya coefficient and the KL divergence.  The ROC curve, BC, and $d_{KL}$ were all estimated based on Monte Carlo with $10^7$ samples.
  \begin{figure}[thbp]
    \centering
    \includegraphics[scale=0.7]{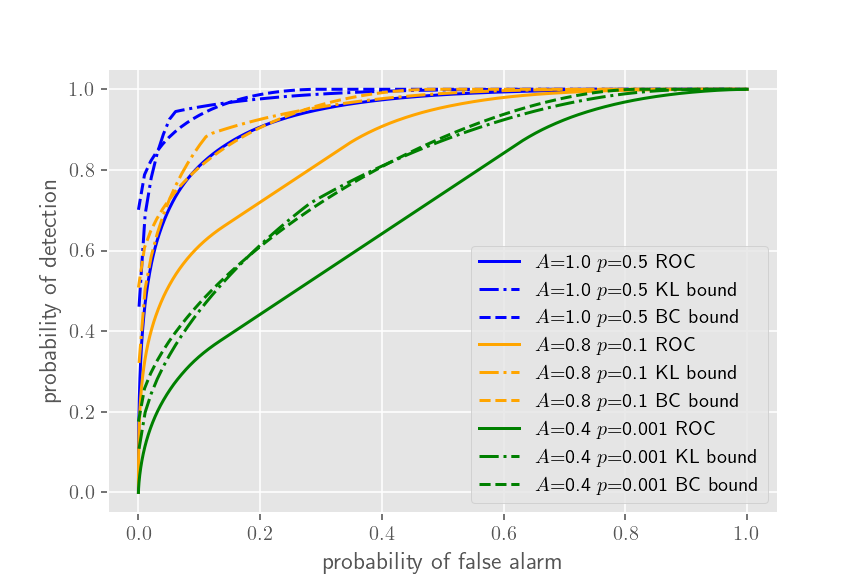}  
    \caption{ROC and bounds for Markov signal with $n=10$.}
    \label{fig:markovROC}
  \end{figure}
  The two bounds are close to each other and they are less tight than in the example with $\|X\|\equiv \mu.$  The distribution of $\|X\|$ is more skewed for smaller $p$.  The linear portions of slope one in the ROC curves in Fig.~\ref{fig:markovROC} are due to the fact the likelihood ratio is one when $\|X\|=0.$  Note in particular that $p_{fa} + (1 - p_{det}) \geq P\{\|X\| = 0\} \geq \frac{1-np} 2$ or $p_{det} \leq p_{fa} + \frac{1 + np} 2$ for all $A>0.$
\end{exam}


\begin{thebibliography}{10}
\providecommand{\url}[1]{#1}
\csname url@samestyle\endcsname
\providecommand{\newblock}{\relax}
\providecommand{\bibinfo}[2]{#2}
\providecommand{\BIBentrySTDinterwordspacing}{\spaceskip=0pt\relax}
\providecommand{\BIBentryALTinterwordstretchfactor}{4}
\providecommand{\BIBentryALTinterwordspacing}{\spaceskip=\fontdimen2\font plus
\BIBentryALTinterwordstretchfactor\fontdimen3\font minus \fontdimen4\font\relax}
\providecommand{\BIBforeignlanguage}[2]{{%
\expandafter\ifx\csname l@#1\endcsname\relax
\typeout{** WARNING: IEEEtran.bst: No hyphenation pattern has been}%
\typeout{** loaded for the language `#1'. Using the pattern for}%
\typeout{** the default language instead.}%
\else
\language=\csname l@#1\endcsname
\fi
#2}}
\providecommand{\BIBdecl}{\relax}
\BIBdecl

\bibitem{Alon06}
U.~Alon, \emph{An Introduction to Systems Biology: {D}esign Principles of Biological Circuits}.\hskip 1em plus 0.5em minus 0.4em\relax CRC press, 2006.

\bibitem{MarbachPrillSchaffter10}
D.~Marbach, R.~J. Prill, T.~Schaffter, C.~Mattiussi, D.~Floreano, and G.~Stolovitzky, ``Revealing strengths and weaknesses of methods for gene network inference,'' \emph{Proc Natl Acad Sci USA}, vol. 107, no.~14, pp. 6286--6291, 2010.

\bibitem{MarbachCostelloKuffner12}
D.~Marbach, J.~C. Costello, R.~K{\"u}ffner, N.~M. Vega, R.~J. Prill, D.~M. Camacho, K.~R. Allison, M.~Kellis, J.~J. Collins, and G.~Stolovitzky, ``Wisdom of crowds for robust gene network inference,'' \emph{Nat Methods}, vol.~9, no.~8, pp. 796--804, Jul. 2012.

\bibitem{SunTaylorBollt15}
J.~Sun, D.~Taylor, and E.~M. Bollt, ``Causal network inference by optimal causation entropy,'' \emph{{SIAM} Journal on Applied Dynamical Systems}, vol.~14, no.~1, pp. 73--106, Jan. 2015.

\bibitem{HajekKang25}
B.~Hajek and X.~Kang, ``Maximum likelihood estimation of optimal receiver operating characteristic curves from likelihood ratio observations,'' \emph{arXiv preprint arXiv:2202.01956}, June 2025.

\bibitem{SimchowitzManiaTu18}
\BIBentryALTinterwordspacing
M.~Simchowitz, H.~Mania, S.~Tu, M.~I. Jordan, and B.~Recht, ``Learning without mixing: Towards a sharp analysis of linear system identification,'' \emph{CoRR}, vol. abs/1802.08334, 2018. [Online]. Available: \url{http://arxiv.org/abs/1802.08334}
\BIBentrySTDinterwordspacing

\bibitem{FattahiSojoudi18a}
\BIBentryALTinterwordspacing
S.~Fattahi and S.~Sojoudi, ``Sample complexity of sparse system identification problem,'' \emph{CoRR}, vol. abs/1803.07753, 2018. [Online]. Available: \url{http://arxiv.org/abs/1803.07753}
\BIBentrySTDinterwordspacing

\bibitem{FattahiMatniSojoudi19}
\BIBentryALTinterwordspacing
S.~Fattahi, N.~Matni, and S.~Sojoudi, ``Learning sparse dynamical systems from a single sample trajectory,'' \emph{CoRR}, vol. abs/1904.09396, 2019. [Online]. Available: \url{http://arxiv.org/abs/1904.09396}
\BIBentrySTDinterwordspacing

\bibitem{OymakOzay19}
S.~Oymak and N.~Ozay, ``Non-asymptotic identification of {LTI} systems from a single trajectory,'' in \emph{2019 American Control Conference ({ACC})}.\hskip 1em plus 0.5em minus 0.4em\relax {IEEE}, Jul. 2019, pp. 5655--5661.

\bibitem{LaleAzizzadenesheliHassibi20}
\BIBentryALTinterwordspacing
S.~Lale, K.~Azizzadenesheli, B.~Hassibi, and A.~Anandkumar, ``Logarithmic regret bound in partially observable linear dynamical systems,'' \emph{CoRR}, vol. abs/2003.11227, 2020. [Online]. Available: \url{https://arxiv.org/abs/2003.11227}
\BIBentrySTDinterwordspacing

\bibitem{BentoIbrahimiMontanari10}
\BIBentryALTinterwordspacing
J.~Bento, M.~Ibrahimi, and A.~Montanari, ``Learning networks of stochastic differential equations,'' in \emph{Advances in Neural Information Processing Systems (NIPS)}, 2010, pp. 172--180. [Online]. Available: \url{http://papers.nips.cc/paper/4055-learning-networks-of-stochastic-differential-equations.pdf}
\BIBentrySTDinterwordspacing

\bibitem{Fuchs05}
J.~Fuchs, ``Recovery of exact sparse representations in the presence of bounded noise,'' \emph{{IEEE} Transactions on Information Theory}, vol.~51, no.~10, pp. 3601--3608, Oct. 2005.

\bibitem{Tropp06}
J.~Tropp, ``Just relax: convex programming methods for identifying sparse signals in noise,'' \emph{{IEEE} Transactions on Information Theory}, vol.~52, no.~3, pp. 1030--1051, Mar. 2006.

\bibitem{ZhaoYu06}
P.~Zhao and B.~Yu, ``On model selection consistency of {Lasso},'' \emph{J Mach Learn Res}, vol.~7, pp. 2541--2563, Nov. 2006.

\bibitem{MeinshausenBuhlmann06}
N.~Meinshausen and P.~B{\"u}hlmann, ``High-dimensional graphs and variable selection with the lasso,'' \emph{The Annals of Statistics}, vol.~34, no.~3, pp. 1436--1462, Jun. 2006.

\bibitem{Wainwright09}
M.~J. Wainwright, ``Sharp thresholds for high-dimensional and noisy sparsity recovery using $\ell_1$-constrained quadratic programming ({Lasso}),'' \emph{{IEEE} Transactions on Information Theory}, vol.~55, no.~5, pp. 2183--2202, May 2009.

\bibitem{JedraProutiere19}
Y.~Jedra and A.~Prouti\`{e}re, ``Sample complexity lower bounds for linear system identification,'' in \emph{2019 {IEEE} 58th Conference on Decision and Control ({CDC})}.\hskip 1em plus 0.5em minus 0.4em\relax {IEEE}, Dec. 2019, pp. 2676--2681.

\bibitem{BentoIbrahimiMontanari11}
J.~Bento, M.~Ibrahimi, and A.~Montanari, ``Information theoretic limits on learning stochastic differential equations,'' in \emph{2011 {IEEE} International Symposium on Information Theory Proceedings}.\hskip 1em plus 0.5em minus 0.4em\relax {IEEE}, Jul. 2011, pp. 855--859.

\bibitem{BentoIbrahimi14}
J.~Bento and M.~Ibrahimi, ``Support recovery for the drift coefficient of high-dimensional diffusions,'' \emph{{IEEE} Transactions on Information Theory}, vol.~60, no.~7, pp. 4026--4049, Jul. 2014.

\bibitem{Kang21}
\BIBentryALTinterwordspacing
X.~Kang, ``Causal network inference simulations,'' Feb. 2021. [Online]. Available: \url{https://github.com/Veggente/net-inf-eval}
\BIBentrySTDinterwordspacing

\bibitem{JedraProutiere23}
Y.~Jedra and A.~Prouti\`{e}re, ``Finite-time identification of linear systems: Fundamental limits and optimal algorithms,'' \emph{IEEE Transactions on Automatic Control}, no.~5, pp. 2805 -- 2820, May 2023.

\bibitem{PedregosaVaroquauxGramfort11}
F.~Pedregosa, G.~Varoquaux, A.~Gramfort, V.~Michel, B.~Thirion, O.~Grisel, M.~Blondel, P.~Prettenhofer, R.~Weiss, V.~Dubourg, J.~Vanderplas, A.~Passos, D.~Cournapeau, M.~Brucher, M.~Perrot, and E.~Duchesnay, ``Scikit-learn: Machine learning in {P}ython,'' \emph{Journal of Machine Learning Research}, vol.~12, pp. 2825--2830, 2011.

\bibitem{CoverThomas06}
T.~Cover and J.~Thomas, \emph{Elements of Information Theory}, 2nd~ed.\hskip 1em plus 0.5em minus 0.4em\relax Wiley-Interscience, 2006.

\bibitem{KangHajekHanzawa20}
X.~Kang, B.~Hajek, and Y.~Hanzawa, ``From graph topology to {ODE} models for gene regulatory networks,'' \emph{{PLOS} {ONE}}, vol.~15, no.~6, p. e0235070, Jun. 2020.

\bibitem{Csiszar67}
I.~Csisz\'{a}r, ``Information-type measures of different probability distributions and indirect observations,'' \emph{Studia Sci. Math. Hungar.}, vol.~2, pp. 299--318, 1967.

\bibitem{Polyanskiy_Wu_2025}
Y.~Polyanskiy and Y.~Wu, \emph{Information Theory: From Coding to Learning}.\hskip 1em plus 0.5em minus 0.4em\relax Cambridge University Press, 2025.

\end{thebibliography}
\end{document}